\documentstyle[prl,aps]{revtex}
\twocolumn
\begin{document}
\input{epsf}
\draft
\twocolumn[\hsize\textwidth\columnwidth\hsize\csname@twocolumnfalse\endcsname
\title{Vorticity Generation in Slow Cooling Flows}
\author{Ami Glasner, Eli Livne, and Baruch Meerson}
\address{The Racah Institute  of  Physics, the Hebrew
University   of  Jerusalem,
Jerusalem 91904, Israel}
\maketitle
\begin{abstract}
We show that any generic non-adiabatic slow flow of ideal 
compressible fluid
develops a significant vorticity.
As an example, an initially irrotational conductive cooling 
flow (CF) is considered. A perturbation theory 
for the vorticity 
generation is developed that employs, as a 
zero order solution, a novel two-dimensional 
similarity solution. 
Full gasdynamic simulations of this CF demonstrate the
vorticity generation and support the theory. The relevance 
of this problem
to the experiments with the ``hot channels" is discussed. 
\end{abstract}
\vskip1pc]
\narrowtext

The most general mechanism of the vorticity 
production in non-adiabatic flows of ideal compressible fluids 
relies on the misalignment of pressure and density gradients \cite{Pedlosky}.
Recent experiments with
supersonic flows \cite{Grun,Jacobs}  have clearly demonstrated the 
efficiency of this mechanism (which is called baroclinic). 
The baroclinic mechanism 
can also
operate in slow gas flows, and its specific manifestations 
in meteorology \cite{Pedlosky} and combustion \cite{Miyauchi} 
are known. The main objective of 
this Letter is to show that the vorticity production 
represents a generic and significant
property of {\it any} slow non-adiabatic gas flow.

For concreteness, we will consider the conductive cooling flows (CFs)
and refer to the ``hot channels" produced in 
the air by 
lasers or electric 
discharges \cite{Greig1,Greig2}. After
pressure equilibration these channels 
develop a significant vorticity and small scale 
turbulence and
cool much faster 
than they would because of molecular thermal conduction. Picone and 
Boris \cite{Picone} interpreted these results 
in terms of
the 
baroclinic vorticity production {\em during the rapid channel expansion} 
(that is, on the acoustic time scale) \cite{Hill1}.  Schlieren 
photographs of the hot channels \cite{Greig1,Greig2}
clearly show 
that 
the most significant vorticity dynamics occurs on a much longer 
time scale. According to
Picone and Boris, 
"after pressure equilibration...  
vorticity is no longer generated, however, significant residual 
vorticity exists" \cite{Picone}. We wish to present an alternative 
scenario which assumes that a significant
vorticity is {\em created} on the long, heat-conduction time scale. 
The underlying physics 
is the following. After a few acoustic times, following the rapid 
energy release,
the gas 
pressure becomes very close to the (constant) ambient pressure, 
while the vorticity
generated earlier is presumably damped out. As the 
temperature inside 
the channel is
still very high, a low-Mach-number conductive CF 
develops that cools
the channel by filling it with the cold gas from the 
periphery. Slow conductive CFs were studied previously in the 
context of a ``point-like"
energy release, like a high-altitude explosion in the atmosphere 
\cite{Meerson} or
``fireball" produced by a laser spark in 
front of condensed matter \cite{Kaganovich}.
We will show that, 
unless 
the energy release geometry is {\it fully}
symmetrical, small pressure gradients, intrinsic in the CF, 
result in a significant vorticity production. 
 
The simplest conductive CF of a perfect gas is described by 
the standard gasdynamic equations \cite{Landau}: 

\begin{equation}
\frac{d \rho}{d t} + \rho \nabla \cdot {\bf v} = 0,
\label{1}
\end{equation}

\begin{equation}
\epsilon^2 \rho \frac{d{\bf v}}{dt} = 
-\nabla p ,\label{2}
\end{equation}

\begin{equation}
\gamma^{-1} \frac{dp}{dt} + p \nabla \cdot {\bf v} - 
\nabla \cdot (T^{\nu} \nabla T)= 0, \label{3}
\end{equation}
where $d/dt= \partial/\partial t + {\bf v} \cdot \nabla$ is the 
total derivative. The distance is measured in the units
of a characteristic spatial scale of the problem $r_0$ (see later), 
while 
the time is measured in the units of the heat conduction time
$\tau_0=\gamma (\gamma - 1)^{-1}R_g \rho_0 r_0^2 \kappa^{-1} T_0^{-\nu}$. 
Furthermore, the gas density $\rho$ and temperature $T$ are scaled by 
their (presumably constant) 
values "at infinity" $\rho_0$ and $T_0$, the velocity $v$ is scaled by
$r_0/\tau_0$, and the pressure $p=\rho T$ is scaled by $(R_g/\mu) \rho_0 T_0$. 
The 
non-dimensional parameter $\epsilon=r_0/c_s\tau_0$ (where $c_s^2=R_g T_0/\mu$)
represents the characteristic 
Mach number 
of the flow. Finally, $R_g, \gamma$ and 
$\mu$
are the gas constant,
adiabatic index and molar mass, respectively, 
while the heat conductivity is assumed
to be a power-like function of the temperature: $\kappa T^{\nu}$ in 
the scaled units, $\kappa=const$. (For the 
molecular air $\nu=1/2$.)

We start with a perturbation theory that describes 
the initial stage of
the vorticity 
production. Then we report on numerical simulations with
the full equations (\ref{1})-(\ref{3}) that show the vorticity generation
in the same CF and support the theory. 
 
In the low Mach 
number regime, $\epsilon^2 \ll 1$, the temperature and
density contrasts can still be large, but
pressure non-uniformities are already small: $ p({\bf r},t) = 1 + 
\epsilon^2 \, \delta p ({\bf r},t)$. 
Then, neglecting the
small $\delta p$ terms in Eq. (\ref{3}) and
equation of state, we obtain 
$ \nabla \cdot ({\bf v} - T^{\nu} \nabla T) = 0$ and 
$\rho T =1$, respectively. It follows that ${\bf v} = {\bf v_p}+ {\bf v_s}$,
where ${\bf v_p}=  - \rho^{-\nu - 2} \nabla \rho$ is the 
irrotational component of the fluid velocity, and ${\bf v}_s$ is 
the solenoidal component: $\nabla \cdot {\bf v}_s = 0$.
Substitution of ${\bf v}$ 
into Eq. (\ref{1}) yields a nonlinear transport equation: 
\begin{equation}
\frac{\partial \rho}{\partial t} + ({\bf v}_s\cdot \nabla) \rho =
\nabla \cdot (\rho^{-\nu-1}\nabla \rho).
\label{9}
\end{equation}
An important additional equation 
follows from Eq. (\ref{2}):
\begin{equation}
\frac{\partial \vec{\omega}}{\partial t} - \nabla \times ({\bf v} 
\times \vec{\omega}) = 
\frac{d{\bf v}}{dt} \times \frac{\nabla \rho}{\rho},
\label{10}
\end{equation}
where $\vec{\omega} = \nabla \times {\bf v} \equiv 
\nabla \times {\bf v_s}$ is the vorticity. Eq. (\ref{10}) 
is equivalent to the well-known vorticity equation \cite{Pedlosky}, 
as its right-hand-side
can be rewritten as $\nabla \rho \times\nabla \delta p/\rho^2$. Note that 
Eq. 
(\ref{10}) does not include $\epsilon$, 
therefore, the vorticity production rate is, in general, of order unity.

In this Letter we 
address the vorticity production in an initially
curl-free flow. Accordingly, we assume 
that $v_s \ll v_p$ and, in the zero order, neglect 
the second term
in the left side of Eq. (\ref{9}). The resulting nonlinear diffusion 
equation,
\begin{equation}
\frac{\partial \rho}{\partial t} =
\nabla \cdot (\rho^{-\nu-1}\nabla \rho),
\label{12}
\end{equation}
describes such a curl-free CF completely \cite{Meerson,Kaganovich}. Now we 
consider a 
first-order 
version of Eq. (\ref{10}), rewritten in terms of 
the vector field ${\bf a} ({\bf r}, t) = \partial {\bf v_s}/\partial t$:
\begin{equation}
\nabla \times {\bf a} + \frac{\nabla \rho}{\rho} \times {\bf a} =
\left[ \frac{ \partial {\bf v_p} }{\partial t} + 
({\bf v_p} \cdot \nabla){\bf v_p} \right] \times \frac{\nabla \rho}{\rho},
\label{14}
\end{equation}
with $\rho$ and ${\bf v_p}$ given by the curl-free 
solution \cite{applicability}. 
Again, Eq. (\ref{14}) shows that for a generic CF 
the vorticity production rate is of order unity. Therefore, 
the
solenoidal part of the velocity field finally becomes comparable to its
irrotational part (at which stage this perturbation scheme breaks down). 

Let us concentrate on a two-dimensional (2d) flow 
in the $xy$-plane with no $z$-dependence, where one can produce 
the first two ``classes of asymmetries" of the hot 
channels \cite{Picone}: (i) off-center laser beam propagation and
(ii) non-circular cross section of the beam. Introduce a modified 
stream function $\psi(x,y,t)$, so that 
$a_x = -\partial \psi/\partial y$ and 
$a_y = \partial \psi/\partial x$. Eq. (\ref{14}) becomes a scalar 
equation for $\psi$:
\begin{equation}
\nabla \cdot (\rho \nabla \psi) =
\left[   
\left( \frac{\partial{\bf v_p}}{\partial t} + 
({\bf v_p} \cdot \nabla){\bf v_p}\right) \times
\nabla \rho \right] \cdot {\bf e_z},
\label{15}
\end{equation}
where ${\bf e_z}$ is the unit vector in the $z$-direction.

One should, however, deal first with Eq. (\ref{12}) and
find the zero-order solutions $\rho(x,y,t)$ and ${\bf v_p}(x,y,t)$ entering
Eq. (\ref{15}). Remarkably, Eq. (\ref{12}) has a family of 
2d-similarity solutions of the second kind:
\begin{equation}
\rho (x,y,t) = t^{\frac{1-2\beta}{1+\nu}} 
R(\xi, \eta),
\label{16}
\end{equation}
where $\xi=x/t^{\beta}$, $\eta=y/t^{\beta}$, and $\beta$ is an arbitrary real 
parameter \cite{3d-a}. Selection 
of parameter
$\beta$ requires the use of initial or boundary conditions. We
shall adopt the following initial density profile: 
$\rho(x,y, t=0) = A^{-1} r^k f(\phi)$, where $r$ and $\phi$ are the polar 
coordinates in the plane $xy$, and $A$ and $k$ are constants. In the case of a 
cylindrically-symmetric 
[$f(\phi)=1$] explosion along the $z$-axis, this profile
with $k=2/(\gamma-1)$ represents 
the $r\rightarrow 0$ density asymptotics 
that sets in at the
end of the expansion stage \cite{Zel'dovich,Korobeinikov}. The function
$f(\phi)$ describes asymmetry. In analogy to 
Ref. \onlinecite{Meerson}, we extend this initial 
condition to the whole CF region. This idealization is justified 
(see Refs. \onlinecite{Kaganovich,Meerson} and gasdynamic simulations 
below) as long
as the density (temperature) contrast in the system remains much 
larger than unity. 

The initial condition yields
$\beta=(\nu k + k +2)^{-1}$. 
Using
Eq. (\ref{16}), one arrives at a 
nonlinear elliptic
equation for the shape function $R(\xi, \eta)$:
\begin{eqnarray}
\frac{\partial}{\partial \xi} \left(R^{-1-\nu} 
\frac{\partial R}{\partial \xi} \right) + \frac{\partial}{\partial \eta} 
\left(R^{-1-\nu} 
\frac{\partial R}{\partial \eta} \right) 
\nonumber\\
+ (\nu k + k + 2)^{-1} \left(\xi \frac{\partial R}
{\partial \xi} +
\eta \frac{\partial R}{\partial \eta} - k R \right) = 0
\label{20}
\end{eqnarray}
(we got rid of the constant $A$ by choosing $r_0=A^{1/k}$). 
We assume for 
simplicity that the initial density profile 
[and, hence, $R(\xi, \eta)$] is symmetric
with respect 
to each of the Cartesian axes and 
solve Eq. (\ref{20}) in the first quadrant 
with the no-flux boundary conditions at the
$\xi$- and 
$\eta$-axes. In addition, we must require that
$R(\xi \rightarrow \infty, \eta \rightarrow \infty) = 
\hat{r}^k f(\hat{\phi})$,
where $\hat{r}$ and $\hat{\phi}$ are the polar coordinates in the plane
$\xi,\eta$. Fig. 1 shows $R(\xi,\eta)$ found 
numerically in a finite square for 
$f(\hat{\phi}) =  1+a \, \cos 2 \hat{\phi}$. We took the usual values
$\gamma=1.4$ and $\nu=0.5$ for the molecular air, and chose $a=0.6$. 
(In this case $\beta=2/19$, while the 
gas density at the channel axis grows in time like $t^{10/19}$.)

\begin{figure}[h]
\vspace{1.0cm}
\hspace{-3.5cm}
\rightline{ \epsfxsize = 5.0cm \epsffile{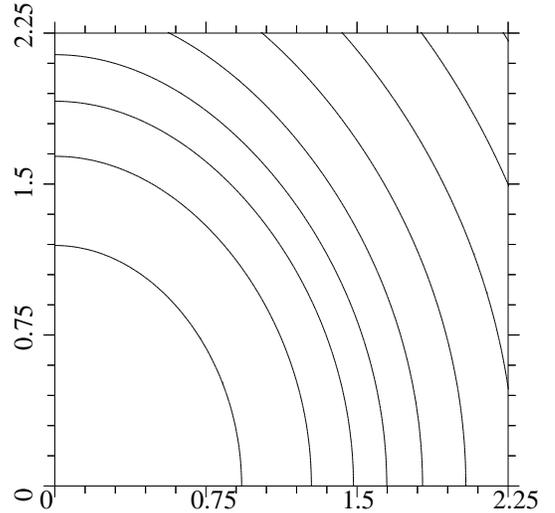}}
\caption{
Contours of $\lg R (\xi,\eta)$. 
On the inner line $\lg R=0.25$
and increases outwards in steps of 0.25. Also, $R(0,0)\approx 1.99$.
\label{Fig. 1}}
\end{figure}

Now we return to Eq. (\ref{15}). 
A similarity solution for $\rho$ implies a similarity 
solution for $\psi$, that is $\psi (x,y,t) = t^{-\alpha} \Psi(\xi,\eta)$,
where $\alpha=2(\nu k + k + 1)/(\nu k + k + 2)$. For the shape 
function $\Psi(\xi,\eta)$ one obtains a linear 
elliptic equation $\nabla \cdot (R \nabla \Psi) =
({\bf W} \times \nabla R) \cdot {\bf e_z}$, where 
\begin{eqnarray}
{\bf W} = -(\nu k + k + 2)^{-1}\left[(\nu k + k + 1){\bf V} + 
\xi \frac{\partial {\bf V}}{\partial \xi} + 
\eta \frac{\partial {\bf V}}{\partial \eta}\right] 
\nonumber\\ 
+ ({\bf V} \cdot \nabla){\bf V}\,,
\nonumber
\end{eqnarray}
${\bf V} = - R^{-\nu-2} \nabla R$, and the $\nabla$-operator now involves
differentiation with respect to $\xi$ and $\eta$. This equation should
be solved in the first quadrant with the Dirichlet boundary condition. 
We solved it 
numerically using the shape function $R$ found earlier (in this case 
$\alpha = 34/19 \approx 1.79$). The result is shown in Fig. 2. 
Now we can evaluate the vorticity $\vec {\omega}= 
\omega(x,y,t)\, {\bf e_z}$,
using the relation $(\partial \omega/\partial t)_{x,y}=\nabla^2 \psi$. 
Following Picone and Boris \cite{Picone}, we
calculate the vorticity flux $\Omega$ 
through the first
quadrant as a function of time. The growth rate of this quantity,
$d\Omega/dt$, is equal to
\begin{equation}
\int_0^{\infty}\int_0^{\infty}dx dy\, \nabla^2 \psi=
\oint_C(\nabla \psi \cdot {\bf n})\, dl,
\label{24}
\end{equation}
where $C$ is the contour going from infinity to zero along the $y$-axis and
continuing to infinity along the $x$-axis and ${\bf n}$ is the 
external normal. Employing the similarity 
solution for $\psi$, we obtain
\begin{equation}
\frac{d \Omega}{d t}= 
-(t+t_0)^{-\alpha}
\left[\int_0^{\infty}\frac{\partial \Psi}{\partial \xi} (0, \eta)\, d\eta
+ \int_0^{\infty}\frac{\partial \Psi}{\partial \eta} (\xi, 0)\, 
d\xi \right],
\label{25}
\end{equation}
where we have
used the invariance of the similarity solution with respect to a time shift 
and introduced $t_0$, the only fitting parameter of the theory. 
Integrating 
Eq. (\ref{25}) with a zero initial 
condition, we arrive at
\begin{equation}
\mid \Omega(t)\mid=
\mid B \mid(\alpha-1)^{-1}\left[t_0^{1-\alpha}-(t+t_0)^{1-\alpha}\right],
\label{26}
\end{equation}
where $B$ is the constant given by the expression in the square 
brackets in Eq. (\ref{25}).  
Eq. (\ref{26}) predicts a linear growth 
of $\Omega$
with time followed by saturation at a constant value 
$\mid B \mid (\alpha-1)^{-1}t_0^{1-\alpha}$. 

\begin{figure}[h]
\vspace{1cm}
\hspace{-4.0cm}
\rightline{ \epsfxsize = 4.0cm \epsffile{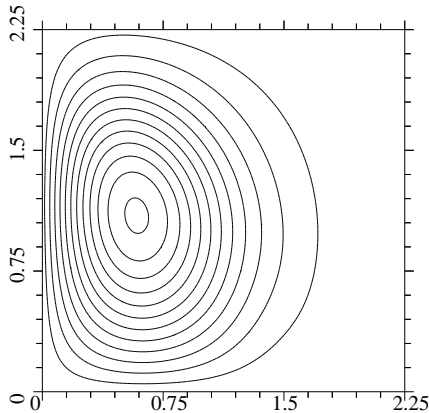}}
\caption{
Contours of $\Psi(\xi,\eta)$. 
On the outer 
line $\Psi=10^{-4}$ and increases inwards in steps of $10^{-4}$.
\label{Fig. 2}}
\end{figure}

Now we report on the 2d-numerical simulations 
with Eqs. (\ref{1})-(\ref{3}). We employed 
an extended version of the 
code VULCAN 
\cite{liv93} that uses flexible moving grids 
and can operate in 
any combination of Eulerian and Lagrangian modes. In the rezoning stage we
used the scheme of Van Leer \cite{Van79} that preserves second order 
accuracy. The code could work in an implicit mode, thus 
eliminating the Courant-Friedrichs-Lewy restriction on the time step.  
The initial conditions were
\begin{equation}
\rho(r,\phi,t=0) \equiv \rho_{in} =\frac{\delta+r^k (1+a \cos 2 \phi)}
 {1+r^k (1+a \cos 2 \phi)}  
\,,
\label{27}
\end{equation} 
for the density,
${\bf v}(r,\phi,t=0)= -\rho_{in}^{-\nu-2} \nabla \rho_{in}$ for the  
velocity, and unity for the
pressure.
For $\delta \ll 1$, the initial 
density profile 
has an extended part described by $r^k (1+a \cos 2\phi)$ (which yields
the 2d-similarity solution). On the other hand, 
$\rho_{in}$ is non-zero at 
$r=0$ and approaches unity at $r\rightarrow \infty$ as it should. 
In most of simulations
we took  $\delta=10^{-2}$, $a=0.6$ and $\epsilon$ in 
the range of
$10^{-6}$ to $10^{-5}$. Simulations show that
the density history at the channel axis is described very well 
by the
similarity scaling $1.99\,(t+4.2 \times 10^{-5})^{10/19}$ until the 
late stage, 
when the 
density contrast is reduced. 
However, the velocity field 
(that was curl-free in the beginning, Fig. 3a) 
develops a noticeable vorticity 
which spatial structure is similar to that shown in  Fig. 2. Finally, 
a distinctive vortex, advected towards 
the origin by the overdense gas inflow, appears (Fig. 3b). Since the 
problem 
is 
symmetric
with respect to each of the axes, the corresponding 
"full" flow develops four symmetric
vortices.
Fig. 4
shows the time history of the vorticity flux 
through the 
first quadrant, $\Omega$, as found
from the simulations. It is seen that the vorticity 
reaches a significant value. One can also see that the perturbation theory
[Eq. (\ref{26})] underestimates the 
saturated vorticity flux. This is understandable, as the
perturbation scheme fails at large times. Interestingly, the agreement
improves for a smaller value of $t_0$.

In summary, we claim that any generic non-adiabatic 
gas flow 
develops a significant vorticity.  For the low-Mach-number 
conductive CF that we 
have considered in detail, the further vorticity
dynamics (instability?)
is apparently sensitive to geometry (like in the Picone-Boris scenario). 
We did not observe 
turbulence or other significant modification of the bulk
transport properties
in this (still
highly symmetric) 2d-flow. Correspondingly, the ``hot channel" 
riddle requires further investigation. One 
can expect
turbulence to show up in a less symmetric
3d situation,
when perturbations
along the channel axis are introduced.

B.M. acknowledges a valuable discussion with P.V. Sasorov.

\begin{figure}[h]
\vspace{7.0cm}
\hspace{3.0cm}
\rightline{ \epsfxsize = 16.0cm \epsffile{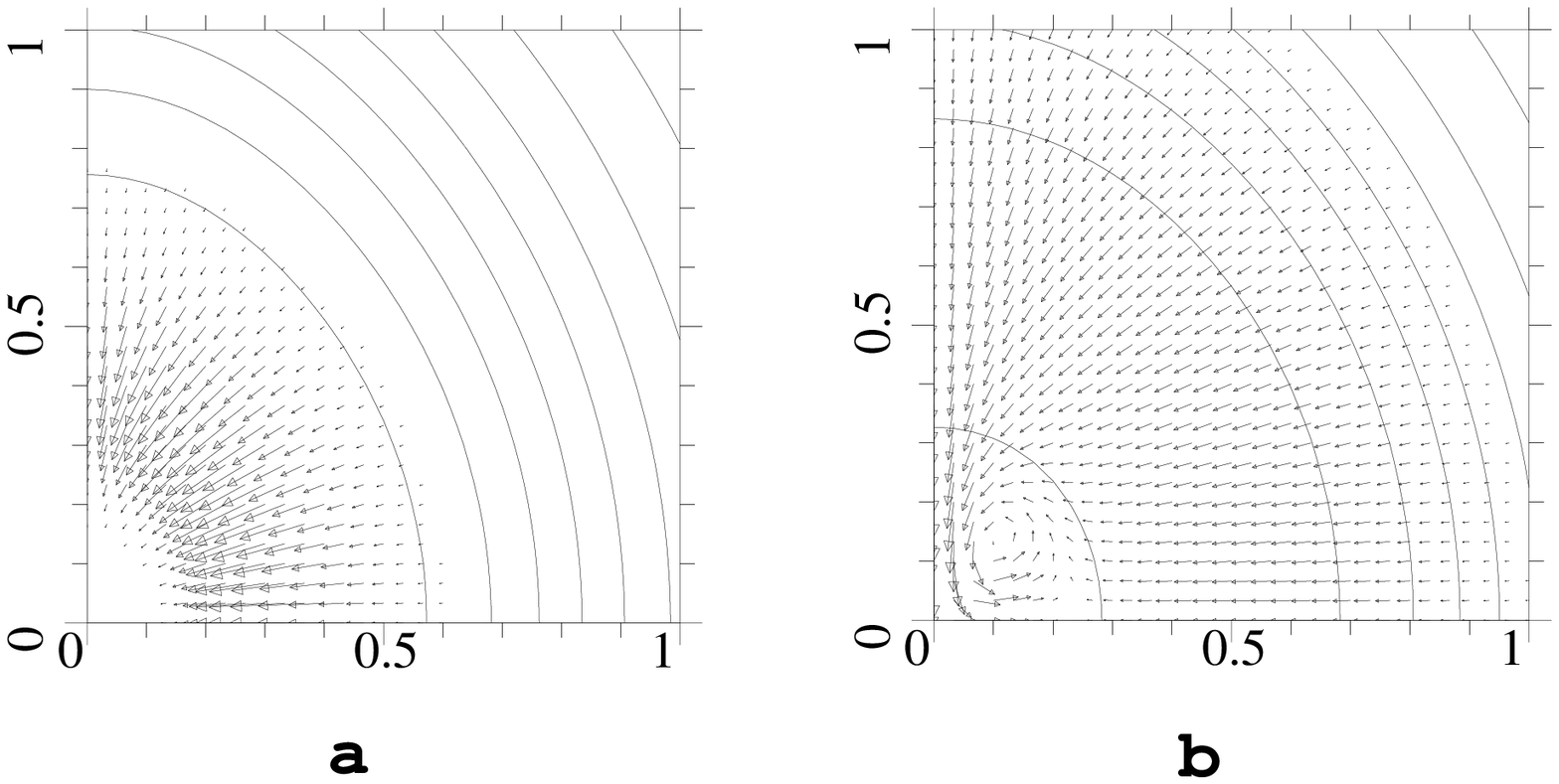}}
\caption{
Density and velocity fields at $t=0$ (a) 
and $t=3 \times 10^{-3}$ (b). The velocity field (arrows) is scaled by 
3000 (a) and 100 (b). On the inner density isolines 
$\rho=0.1$ and increases outwards in steps of 0.1. 
\label{Fig. 3}}
\end{figure}

\begin{figure}[h]
\vspace{1cm}
\hspace{-2.0cm}
\rightline{ \epsfxsize = 6.0cm \epsffile{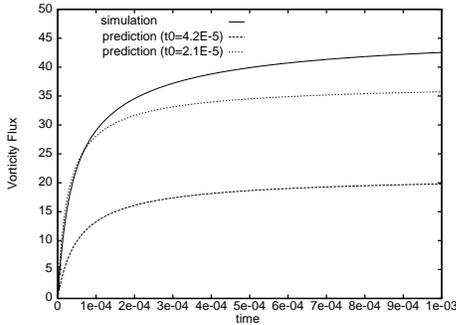}}
\caption{
Vorticity flux through the first quadrant vs time,
as predicted by the full simulations, and by
Eq. (\ref{26}) with different values of $t_0$. 
\label{Fig. 4}}
\end{figure}


\begin{references}
\bibitem{3d-a} In three dimensions, the corresponding
similarity
solutions have an additional argument 
$z/t^{\beta}$.
\end{references}
\end{document}